# ENERGY AND LATENCY AWARE APPLICATION MAPPING ALGORITHM & OPTIMIZATION FOR HOMOGENEOUS 3D NETWORK ON CHIP

Vaibhav Jha[1], Sunny Deol[1], Mohit Jha[2] and G K Sharma[1]

[1]Department of Computer Science & Engineering, Indian Institute of Information Technology and Management, Gwalior, Madhya Pradesh 474015
`vaibhavjha1987@yahoo.com,deol_sunny1986@yahoo.com`
[2]Department of Electrical Engineering, Jabalpur Engineering College, Jabalpur, Madhya Pradesh 482011
`mohitjha_1989@yahoo.com`

*ABSTRACT*

*Energy efficiency is one of the most critical issue in design of System on Chip. In Network On Chip (NoC) based system, energy consumption is influenced dramatically by mapping of Intellectual Property (IP) which affect the performance of the system. In this paper we test the antecedently extant proposed algorithms and introduced a new energy proficient algorithm stand for 3D NoC architecture. In addition a hybrid method has also been implemented using bioinspired optimization (particle swarm optimization) technique. The proposed algorithm has been implemented and evaluated on randomly generated benchmark and real life application such as MMS, Telecom and VOPD. The algorithm has also been tested with the E3S benchmark and has been compared with the existing algorithm (spiral and crinkle) and has shown better reduction in the communication energy consumption and shows improvement in the performance of the system. Comparing our work with spiral and crinkle, experimental result shows that the average reduction in communication energy consumption is 19% with spiral and 17% with crinkle mapping algorithms, while reduction in communication cost is 24% and 21% whereas reduction in latency is of 24% and 22% with spiral and crinkle. Optimizing our work and the existing methods using bio-inspired technique and having the comparison among them an average energy reduction is found to be of 18% and 24%.*

## 1. INTRODUCTION

The scaling of microchip technologies has resulted into large scale Systems-on-Chip (SoC), thus it has now become important to consider interconnection system between the chips. The International Technology Road-map for Semiconductors depicts the on-chip communication issues as the limiting factors for performance and power consumption in current and next generation SoCs [1] [2] [3]. Thus Network on chip has not only come up with an alternative for the SoC, it has also solved the problem faced in the traditional bus based architecture and is an efficient approach towards optimal designs. Although various works has been done in the optimization of the design and the major area where the design need to be focused are Topology, Scheduling*, Mapping, and Routing* [2]. Each area plays an important role in delivering better performance of the system, but in this paper stress is been given on the mapping of the IP core onto the 3d architecture.

By mapping of the IP core we mean, assigning the task in the form of the characterization graph to the given architecture following the design constraint such as area, latency, communication energy and communication cost which should be minimum. As today's application are becoming much more complex their average communication bandwidth are in Gbps and as technology is scaling down, in future a single application would be running on single core thus the bandwidth requirement of link would increase, therefore attention is also given onto its minimization. Figure 1 shows the mapping of the IP core onto 2D mesh architecture. Assigning the given application onto the given architecture is very important from energy point of view, sometime the topology also matters more in optimizing the design [4]. Topology helps in determining latency, traffic pattern, routing algorithm, flow control, and makes huge impact on design cost, power, and performance. Mapping of the IP core onto NoC tile could be done by either assigning a single core onto each tile or by assigning multiple IP core on each of the tile (Scheduling). Each of the mapping procedure has its own merits and demerits. In this we are presenting a similar approach of mapping single IP core on each NoC tile.

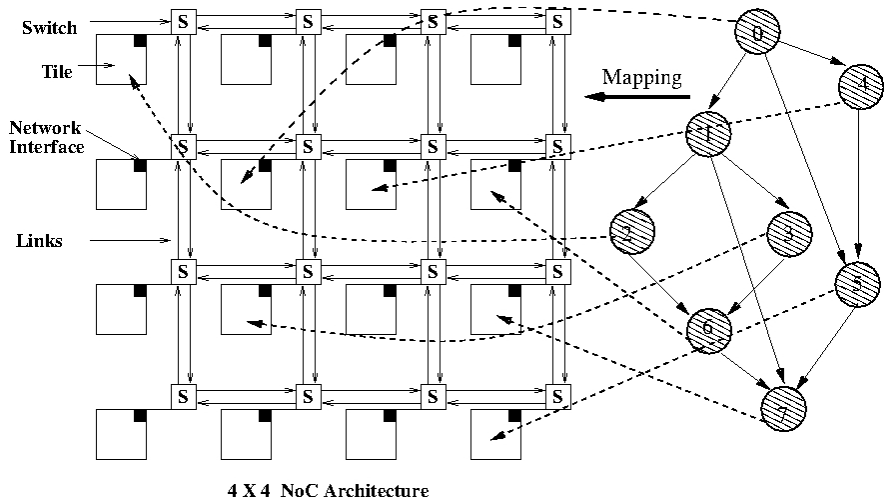

Figure 1. Mapping of cores onto tiles in NoC architecture (2D-Mesh).

In addition [5], include a automated design technique based on Hybrid particle swarm optimization that solve the problem of routing decision with the objective of minimum communication energy consumption. PSO is a technique which randomly places an IP core and tries to minimize the objective function based on swarm intelligence. In this paper hybrid PSO is combined with our proposed algorithm and previously proposed algorithm [6].

## 2. RELATED WORK

Various mapping algorithm has been discussed by the researchers in their literature. *Branch-and-Bound* [7], *Binary Particle Swarm Optimization* (BPSO) [8], based on latency and energy aware, *Diagonal map* [9], are few mapping algorithm which is for 2D NoC architecture but they are also applicable on 3D architecture and shows better results compared to 2D. Task priority based [6], multiple $v_{dd}$ [10] [11] and thermal aware mapping [12], are few work done by the researchers in the field of 3D NoC.

Energy and performance aware mapping using Branch and Bound (PBB) [7] maps the IP core on the basis of the PBB technique in which mapping complexity increases exponentially as the number of the core increases. Another concept of mapping in which the author has focused on latency and energy using the BPSO [8] has proposed mapping as the NP-hard problem and thus the heuristic approach is needed for finding the solution. Author has compared result with the

Genetic algorithm and found BPSO optimal. In D-map [9] procedure, mapping is based on the concept that IP core which is communicating maximum with the rest of the core should be placed on those NoC tile which has greater number of the neighboring tile attached to it thus in 2D architecture diagonal tiles are having maximum number of the neighboring tiles.

In mapping procedure using *multiple $v_{dd}$* Kostas and Dimitrios, [10] [11] has claimed energy of the system could be saved either applying better mapping or the routing procedure or by supplying less voltage to the system and making the design itself optimal. As not each of the router function every time so author divided the whole of the architecture into two layer and each of the layer functioning at different voltage of $v_{dd_{high}}$ and $v_{dd_{low}}$. IP core with greater communication volume are placed onto layer with $v_{dd_{high}}$ and those with low in $v_{dd_{low}}$ when they need to communicate they do with level converter. In his continuing paper [10] author has compared results with different benchmark of MWD, MPEG, VOPD, MMS etc.

In [6] author has proposed two different method of mapping procedure on 3D NoC design *Crinkle* and *Spiral* and task are organized on the task priority list which is based on the basis of the maximum communication volume or maximum out-degree. In [12], author has targeted communication aware used the *Genetic algorithm* for finding out the optimal mapping solution.

In [8], author proposed a heuristic mapping algorithm based on chaotic discrete particle swarm optimization technique. The algorithm resolves the mapping problem to optimize the delay and energy consumption and showing better results than genetic algorithm. In [5], author proposed a routing technique, based on the Hybrid Particle Swarm Optimization (PSO) Algorithm is applied on the 2D-Mesh NoC platform to balance the link load and this algorithm is combined with genetic algorithm, and shown a better results. In [13] author has proposed optimal multi-objective mapping algorithm for NoC architecture.

## 3. DEFINITIONS

Definition 1: An application characterization graph (APCG) G = (C,A) is a directed graph, where each vertex $c_i$ represents selected IP/core, and each directed arc $a_{i,j}$ characterizes the communication from $c_i$ to $c_j$. Each $a_{i,j}$ has application specific information such as:

- $v(a_{i,j})$ is the arc volume from vertex $c_i$ to $c_j$, i.e. the communication volume (bits) from $c_i$ to $c_j$.
- $b(a_{i,j})$ is the arc bandwidth requirement from vertex $c_i$ to $c_j$.

Definition 2: A NoC architecture can be uniquely described by the triple Arch( *T(R, Ch, L), $P_R$ , Ω(C)* ), where:

1) T(R,Ch,L) is directed graph describing the topology. The routers (R), channels (Ch) and the Layers (L) in the network have following attributes:
   a) ∀ *(ch) ϵ Ch, w(ch)* gives the bandwidth of the network channels.
   b) ∀ *(r) ϵ R, I(ch, r)* gives the buffer size(depth) of the *channel ch*, located at *router r*.
   c) ∀ *(r) ϵ R, Pos(r)* specifies the position of *router r* in the floor plan.
   d) ∀ *(l) ϵ L, Layer(l)* specifies the layer of topology.
2) $P_R$(r,s,d) describes the communication paradigm adopted in the network.
   - s,d,r ϵ R, n ⊂ R defines the routing policy at *router r* for all packets with *source s* and *destination d*.
3) Ω *: C → R*, maps each core $c_i$ ϵ C to a router. For direct topologies each router is connected to a core, while in indirect topologies some routers are connected only to other routers.

## 4. PROBLEM FORMULATION

As we are aiming for the minimization of the total communication energy and the total communication energy ($E_{total}$) depends on *Latency, number of hops count, links energy, and switch energy* [7] [13], therefore minimization of each of these factor will result into reduction of global $E_{total}$. Thus our problem has been formulated as:

**Given** an APCG *G(C,A)* and a network topology *T(R, Ch, L)*;
**Find** a mapping function $\Omega : C \rightarrow R$ which maps each core $c_i \in C$ in the APCG to a router $r \in R$ in *T(R, Ch, L)* so that we get :

$$min \left\{ \sum v(a_{i,j}) \times e(r_{map(c_i),map(c_j)}) \right\}$$

**such that:**

$$\forall c_i \in C, map(c_i) \in T$$
$$\forall c_i \neq c_j \in C, map(c_i) \neq map(c_j)$$

where $e(r_{i,j})$ is the average energy consumption of sending 1-bit of data from tile $t_i$ to $t_j$ on which core $c_i$ and $c_j$ are mapped respectively.

Various energy consumption model has been discussed in [7][2], the 1 bit energy consumption in the NoC architecture as proposed in [7] which is calculated as:

$$E_{Bit}^{t_i,t_j} = n_{hops} \times E_{S_{Bit}} + (n_{hops} - 1) \times E_{L_{Bit}} \qquad (1)$$

where, $E_{L_{Bit}}$ is the per bit link energy, $E_{S_{Bit}}$ is per bit switch energy, $E_{Bit}^{t_i,t_j}$ is the total energy when one bit is moved from tile $T_i$ to tile $T_j$, $n_{hops}$ represent the total number of the hops.

Based on below proposed algorithm performance of the system will increased as our approach reduces the number of hops between source and destination. The performance of the system is evaluated on the basis of total communication cost which is calculated as

$$Cost = \sum_{\forall j=1,2,3...|V|, i \neq j} \left\{ b_{a_{i,j}} \times n_{hops}(i,j) \right\}$$

Latency: In a network, latency is a synonym for delay, is an expression of how much time it takes for a packet of data to get from one designated point to another and so that average latency is calculated as:

$$Latency_{avg} = \frac{\sum_{\forall j=1,2,3...|V|, i \neq j} \left\{ n_{hops} \times Comm.Volume \times \rho \right\}}{\eta}$$

where, $\rho$ = constant related to delay
$\eta$ = Total number of times transfer occurred of communication volume between source and destination.

## 5. PROPOSED ALGORITHM

Assignment of an IP core onto the given architecture is called a *mapping*. When we map an IP core onto the given architecture we always try to place those core which is communicating maximum($core_{i,j}^{max}$) closer to each other, limiting the number of hops travelled by data between two related cores, thus the $E_{total}$ gets reduced. As in 2D architecture diagonal tiles(tiles with degree 4) are having greater number of communicating links, so placing $core_i^{max}$ onto these tile and placing rest of unmapped core w.r.t this mapped core onto neighboring tile would reduce the number of hop count thus $E_{total}$ would minimize [9].

Similar approach when carried out with the 3D NoC in which the topology is 2D layered architecture shown in Figure 2, the diagonal tiles of each layer has four adjacent links and one *uplinks* and *downlinks* each. The mapping of highest out-degree core onto diagonal tiles gives the flexibility of assigning highest communicating unmapped cores much closer to mapped cores. This helps to minimize the number of hops travelled by data between two communicating core, which is our aim to minimize the total energy consumption. Question arises how to find the IP core and arrange them such that it could be mapped in regular fashion. In this paper we have used the concept of maximum out degree. We find those cores which is having maximum out degree and order them in an array of descending order, if there are two or more IP cores which is having same number of the out-degree then we differentiate them on the basis of the *Rank* which is calculated as:

$$Ranking\ (Core_i) = \sum_{\forall j=1,2,3\ldots|V|, i \neq j} (comm_{i,j} + comm_{j,i})$$

where $comm_{i,j}$ represents the communication volume(in bits) transferred from $core_i$ to $core_j$. The core to be mapped next is the unmapped core which communicates maximum with the mapped core.

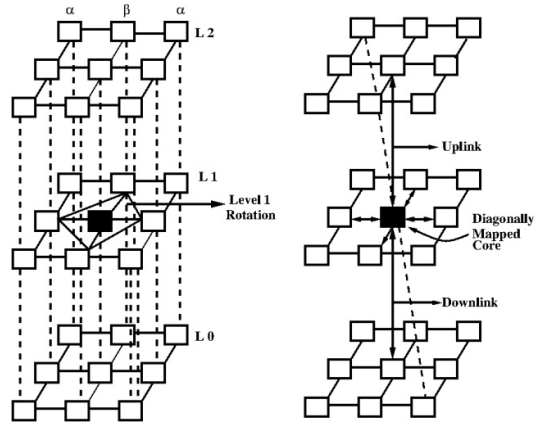

Figure 2. Mapping of highest out-degree cores onto diagonal tiles (shown in dark shaded).

After arranging the IP cores on the basis of the out-degree and rank, the core with maximum out-degree is mapped onto the diagonal tiles. Each layer in the architecture is having only one tile which comes in the list of the diagonal tiles to which these highest out-degree IP core will be mapped. While mapping we leave the diagonal end tiles as it is having at most only three tile to which it can communicate. After mapping the core onto the diagonal tile the position of the next unmapped core is found on the basis of the lozenge shape, in which we apply two rotation namely $\alpha$ and $\beta$ rotation. $\alpha$ rotation works out in clockwise direction for odd numbered column and $\beta$ in anti-clockwise direction for even numbered column discussed in literature [9]. While mapping the IP core onto the each layer we apply this rotation for intra-layer tiles only for finding the position of the next empty tile, but during the mapping process their is possibility that position of the next empty tile could not be there in the same layer so we need to change the layer. While applying the rotation we need to change the level of the rotation, the maximum number of the level which can be changed in each rotation is *2×(n-1)*, where *n* is the dimension of the architecture(*n×n×n*). In calculation of the energy, latency is also important which is dependent on the routing procedure. In our work we have used the *XYZ* routing algorithm for finding out the number of the hops count.

Following the Algorithm 1 the number of hops count between the two related cores is reduced, which in turn reduces the total communication energy consumption given in Equation 1.

**Algorithm 1** Mapping Algorithm for 3D NoC
-----------------------------------------------------------------------------------------------------------------
**Input:** APCG *G(C,A)* and *(n×n×n)* 3D NoC Topology *T(R, Ch, L)*.
**Output:** Array *M[ ]* containing corresponding tile numbers for $c_i \in C$.
1: **Initialize:** *Mapped[ ] = -1;UnMapped[ ] = $c_i \in$ C*
2: **Mapping()** {
3: *Sort ($c_i \in$ C)*
4: *Store in OD[ ]. { in descending order of outdegree & Ranking().}*
5: **for** *i = 0 to n − 3* **do**
6:     *OD[ i ].mapto() → ($n^2 + n + 1$) × (i + 1);*
7:     *Mapped[ i ] = OD[i];*
8:     *Remove OD[ i ] from UnMapped[ ];*
9: **end for**
10: **while** *UnMapped[ ] ≠ empty* **do**
11:     for all $c_i \in$ *UnMapped[ ]*, select $c_i$ where:
12:     *max comm(i,j) { where i ∈ UnMapped[ ] , j ∈ Mapped[ ];}*
13:     *positionof(j); //(row, column, layer) of $c_j$*
14:     *colno = (j.mapto())%n*
15:     **if** *(colno% 2 ≠ 0)* **then**
16:         **while** *(flag = 0)* **do**
17:           *alpharotation()*
18:           **if** *empty tile found* **then**
19:               *return $t_k$, set flag = 1;*
20:           **else**
21:               *layer+ + || layer- - || (both)*
22:               *alpharotaion()*
23:           **end if**
24:         **end while**
25:     **else**
26:         **while** *flag = 0* **do**
27:           *betarotation()*
28:           **if** *empty tile found* **then**
29:               *return $t_k$, set flag = 1*
30:           **else**
31:               *layer+ + || layer- - || (both)*
32:               *betarotation()*
33:           **end if**
34:         **end while** {returns empty tile($t_k$) using Lozenge shape path.}
35:         *OD[ i ].mapto() → $t_k$*
36:         *Mapped[i] = OD[ i ]*
37:         *Remove OD[ i ] from UnMapped[ ]*
38:     **end if**
39: **end while**
40: *calculate_energy() {calculates energy for the mapping generated.}*
-----------------------------------------------------------------------------------------------------------------

# 6. BIO-INSPIRED OPTIMIZATION ALGORITHMS

In this section brief description of the implemented algorithms like Hybrid PSO, Hybrid ARPSO and Hybrid QPSO is discussed.

## 6.1 PARTICLE SWARM OPTIMIZATION (PSO)

Particle Swarm Optimization (PSO) is based on the movement and intelligence of swarms [14]. The fundamental idea behind PSO is the mechanism by which the birds in a flock (swarm) and the fishes in a school (swarm) cooperate while searching for food. Each member of the swarm called particle, represents a potential solution of the problem under consideration. Each particle in the swarm relies on its own experience as well as the experience of its best neighbor. Each particle has an associated fitness value. These particles move through search space with a specified velocity in search of optimal solution. Each particle maintains a memory which helps it in keeping the track of the best position it has achieved so far. This is called the particle0s personal best position *(pbest)* and the best position the swarm has achieved so far is called global best position *(gbest)*. After each iteration, the *pbest* and *gbest* are updated for each particle if a better or more dominating solution (in terms of fitness) is found. This process continues iteratively, until either the desired result is converged upon, or its determined that an acceptable solution cannot be found within computational limits. In search of an optimal solution particles may trap in local optimal solution, therefore some heuristics can be used to help particles get out of local optimum. It is proved that PSO Algorithm 2 for IP mapping has worked much better than various proposed algorithms for routing and task mapping in NoC.

Due to a decrease of diversity (because of clustering of particle) in search space, the PSO tends to suffer from problem of premature convergence which leads to sub optimal solution. The attractive and repulsive PSO (ARPSO) algorithm proposed in [15] overcomes the problem of premature convergence. ARPSO switches between two phases to find an optimal solution:
1) Attraction Phase 2) Repulsion Phase.

In Attraction Phase, particle attracts each other, as in the basic PSO algorithm. In Repulsion Phase, the individual particle is attracted by its own previous best position *(pbest)* and repelled by the global best position *(gbest)*. In this way there is neither total attraction nor total repulsion but a balance between two.

**Algorithm 2** PSO Algorithm for IP Mapping
------------------------------------------------------------------------------------------------------------------
Let *G[ ]* be the current best particle after various simulations.
Let *nb_eval* keeps track on the evaluation number in a simulation.
Let *n_exec* keeps track on the simulation number, it is initialized to 0.
Let *total_cost_min* is the minimum energy achieved in a simulation.
Let *min_cost* is the minimum energy achieved in all simulations executed.
**Input:** APCG, NoC Topology
**Output:** Mapping Sequence
1: *n_exec* ← *n_exec + 1*
2: **for** *s = 0* to *S* **do**
3:       initialize *X[s]* and *V [s]* randomly
4:       *X[s].f* ← *objfn(s)* {evaluating total_cost(objective function) using Equation 1}
5:       *P[s]* ← *X[s]* {local best position of the $s^{th}$ particle}
6: **end for**
7: *P[best]* ← *min_total_cost(P[s])*
8: **while** *nb_eval < eval_max* **do**
9:       **for** *s = 0* to *S* **do**
10:          update *V [s]* and *X[s]* using equation 2 and 3

```
11:         X[s].f ← objfn(S)
12:         if X[s].f < P[s].f then
13:             P[s] ← X[s] {updating local best position of the $s^{th}$ particle}
14:         end if
15:         if X[s].f < P[s].f then
16:             P[s] ← P[best] {updating global best position}
17:         end if
18:     end for
19: total_cost_min ← P[best]
20: end while
21: if total_cost_min < min_cost then
22:     min_cost ← total_cost_min
23:     G[ ] ← P[best]
24: end if
25: if n_exec < n_exec_max then
26:     goto 1
27: end if
```
-------------------------------------------------------------------------------------------------------------

qPSO proposed in [16], another variant of PSO in which authors present the hybridization of PSO with quadratic approximation operator (QA), is implemented to solve IP mapping problem to get an optimal NoC design. The hybridization is performed by splitting the whole swarm into two sub swarms in such a way that the PSO operators are applied on one sub swarm, whereas the QA operator is applied on the other sub swarm, ensuring that both sub swarms are updated using the global best particle of the entire swarm.

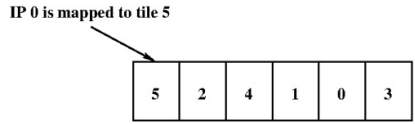

Figure 3. A particle representing a solution.

Basic PSO algorithm is implemented to find the optimal solution for IP mapping problem in NoC designs. Dimension or size (D) of the particle is set equal to the number of tiles in the NoC topology. Firstly, initial population with discrete values is generated having number of particles equal to the swarm size (S) and each particle is initialized with the initial velocity (V) and position (X). Particle for IP mapping is represented as shown in Figure 3, $n^{th}$ IP is mapped on $x_n$, where n is the index in particle. For IP mapping problem objective or fitness function (f) is to minimize *Total Communication Cost* described in Equation (1). Then these particles move into the search space in search of an optimal solution by updating their velocity and position towards its previous best solution (*P[s]*) and the global best solution (*P[best]*) using Equations (2) & (3).

$$v_i^{k+1} = wv_i^k + rand_1(0,c_1) \times (pbest_i - x_i^k) + rand_2(0,c_2) \times (gbest - x_i^k) \qquad (2)$$

and,

$$x_i^{k+1} = x_i^k + \lfloor v_i^{k+1} \rfloor \qquad (3)$$

where,
$\lfloor i \rfloor$ : gives floor value of i
$v_i^k$ : velocity of agent i at iteration k

w : weighting function
$c_1$ and $c_2$ : acceleration coefficients
rand : uniformly distributed random number between 0 and 1
$x_i^k$ : current position of agent i at iteration k

A modification is done in the position updating equation, the floor value of the updated velocity is added to the previous position for getting the new position of the particle, since the search space of IP mapping problem is discrete and the new velocity of particle may be real. For generating the integral position the floor value of velocity is taken. For performing experiments the values of various parameters for PSO are shown in Table 1(based on intensive experiments).

Table 1
PSO PARAMETERS AND THEIR VALUES

| Parameter | Value |
|---|---|
| $c_1$ | 1.2 |
| $c_2$ | 1.3 |
| w | 0.721348 |
| Swarm Size (S) | 200 |
| Dimension of the Search Space (D) | No. of tiles |
| Maximum No. of Simulations | 100 |
| Maximum No. of function evaluations in each simulation | 150000 |

Similarly, ARPSO is also applied to this problem, modification of the particle's position in repulsion phase of ARPSO is mathematically modeled according to the following equations:

$$v_i^{k+1} = wv_i^k + rand_1(0,c_1) \times (pbest_i - x_i^k) - rand_2(0,c_2) \times (gbest - x_i^k) \quad (4)$$

and,

$$x_i^{k+1} = x_i^k + \lfloor v_i^{k+1} \rfloor \quad (5)$$

ARPSO is applied to solve the IP mapping problem in NoC design for all the four benchmarks considered. The procedure to implement ARPSO is same as basic PSO except one parameter i.e. *ator*, which is the number of evaluations after which algorithm enter into repulsion phase from attraction phase. For performing experiments the values of various parameters for ARPSO are shown in Table 2 (based on intensive experiments).

Similarly, qPSO is also applied, the parameter settings for qPSO are shown in Table 3 (based on intensive experiments).

It is worth mentioning that DMAP proposed in [9] is one of the best mapping algorithms in terms of communication energy consumption as it results in a fraction of second. By having the DMAP result and knowing evolutionary nature of PSO algorithm, different mappings with all reasonable ranges of communication energy can be obtained. To do this, DMAP result is injected into population initialization step as a particle in basic PSO algorithm, this result in a new PSO algorithm named HPSO (Hybrid PSO). DMAP result is also injected into population initialization step as a particle in Attractive-Repulsive PSO (ARPSO), qPSO algorithms, this result in a new PSO algorithms named HARPSO, HqPSO shown in Figure 4.

Table 2
ARPSO PARAMETERS AND THEIR VALUES

| Parameter | Value |
|---|---|
| $c_1$ | 1.2 |
| $c_2$ | 1.3 |
| w | 0.721348 |
| ator | 5000 |
| Swarm Size (S) | 200 |
| Dimension of the Search Space (D) | No. of tiles |
| Maximum No. of Simulations | 100 |
| Maximum No. of function evaluations in each simulation | 150000 |

Table 3
QPSO PARAMETERS AND THEIR VALUES

| Parameter | Value |
|---|---|
| $c_1$ | 2.8 |
| $c_2$ | 1.3 |
| w | 0.719 |
| Coefficient of Hybridization (CH) | 30% |
| Swarm Size (S) | 200 |
| Dimension of the Search Space (D) | No. of tiles |
| Maximum No. of Simulations | 100 |
| Maximum No. of function evaluations in each simulation | 150000 |

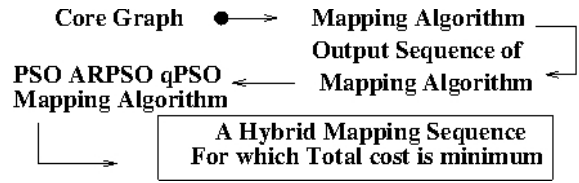

Figure 4. The procedure to achieve optimal mapping solution using HPSO, HARPSO and HQPSO.

## 7. EXPERIMENT RESULTS

In this section we present our experimental results derived from simulations of a combination of different real life applications and E3S benchmark. Our algorithm is compared with two other NoC-targeted mapping algorithms, and results are illustrated and analyzed in details.

Our proposed algorithm has been implemented in C++ and tested with different real life applications (Video Object Plane Decoder (VOPD), Telecom, MMS (Multi-Media System), MWD (Multi-Window Display)), randomly generated benchmarks using TGFF [17] and with E3S benchmark. These tasks are mapped over 3×3×3 3D NoC architecture. We have worked out with XYZ-routing algorithm which provides a deadlock free minimal routing technique. Results obtained are compared with results of various proposed approaches like Spiral, Crinkle which are proposed in literature [6].

Various parameter constraints that have been taken for performing the experiment are shown in Table 4. Number of benchmark over which the system has been tested having a 16 tasks *VOPD*, 16 task *MMS*, 12 tasks *Multi window display*, 16 tasks *Telecom*, 27 task *random* application and E3S(Embedded System Synthesis Benchmarks Suite) application which has 24 tasks *auto industrial*, 12 tasks *consumer*, 13 tasks *networking*, 5 tasks *office automation*.

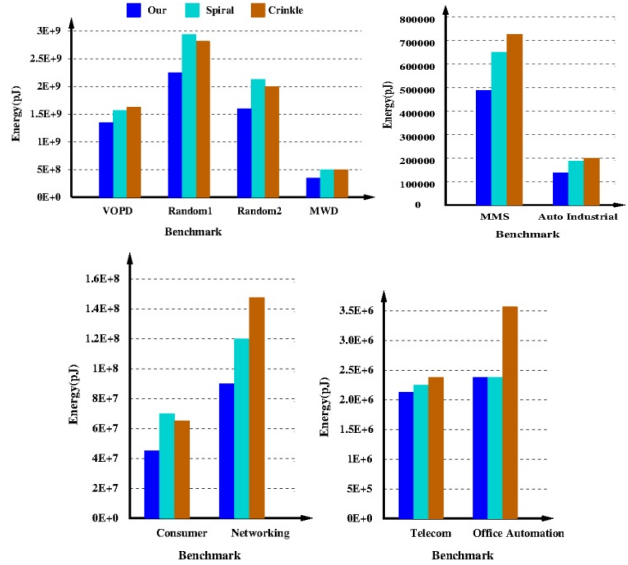

Figure 5. Energy Consumption with different benchmarks using various approaches.

By implementing existing and our proposed algorithms and comparing their result, we found that our algorithm gives optimal result. Testing our proposed algorithm with real life application, random benchmarks and E3S benchmark, we have noticed that our approach give better result (Figure 5) and provide improvement in the communication energy consumption (Figure 9). As discussed above, the average improvement in latency with various benchmarks is shown in Figure 10.

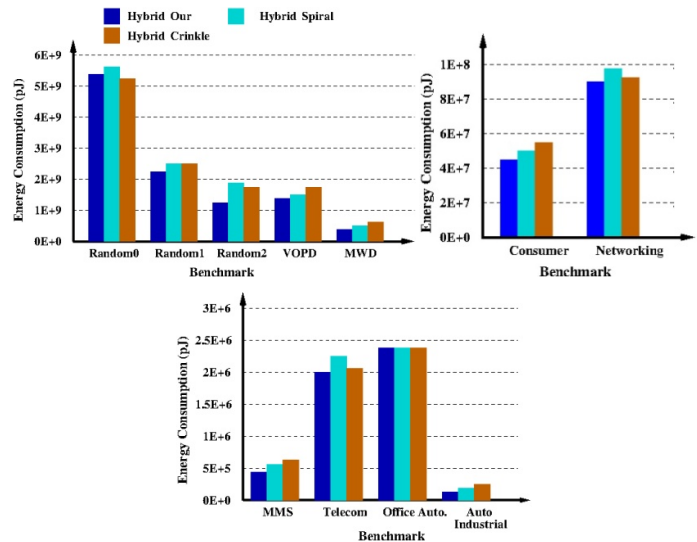

Figure 6. Comparison of Hybridization technique of pso algorithm with Our, Spiral, Crinkle & shows optimal energy consumption results with various benchmarks.

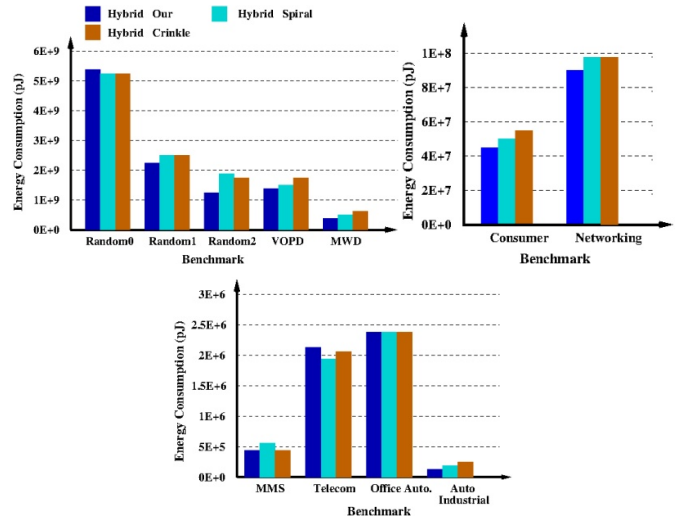

Figure 7. Comparison of Hybridization technique of arpso algorithm with Our, Spiral, Crinkle & shows optimal energy consumption results with various benchmarks.

Implementing various optimization technique like hybrid particle swarm optimization, ARPSO, QPSO and comparing the result of Hybrid PSO+Our approach, Hybrid PSO+Spiral and Hybrid PSO+Crinkle is shown in Figure 6 and the comparison of Hybrid ARPSO+Our approach, Hybrid ARPSO+Spiral and Hybrid ARPSO+Crinkle is shown in Figure 7 and the comparison result for Hybrid QPSO+Our approach, Hybrid QPSO+Spiral and Hybrid QPSO+Crinkle is shown in Figure 8 provide optimal results compare to mapping algorithms. Improvement in the consumption of energy is shown in Figure 11 which is the average of above discussed, all optimization algorithm.

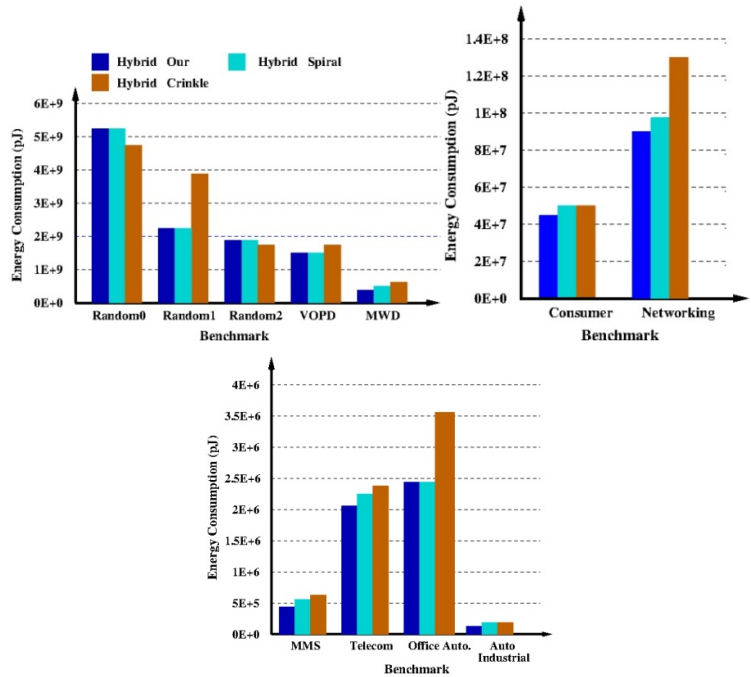

Figure 8. Comparison of Hybridization technique of qpso algorithm with Our, Spiral, Crinkle & shows optimal energy consumption results with various benchmarks.

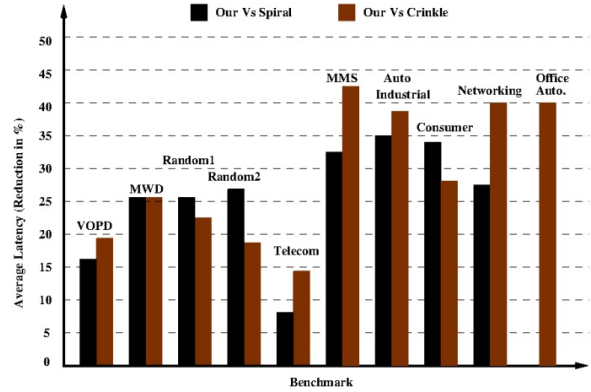

Figure 9. Energy Consumption reduction in % with various benchmarks.

Table 4.
BIT ENERGY VALUES FOR LINK AND SWITCH

| Link | Switch |
|---|---|
| 0.449pJ | 0.284pJ |

## 8. CONCLUSION

In this paper, an efficient energy aware mapping algorithm for 3D NoC has been proposed. As each of the mapping procedure aim at minimizing the number of the hops count between the two communicating core, thus our proposed algorithm also reduced the number of the hops count. On 3D architecture the diagonal tiles of each layer is having greater number of the links so when this algorithm is tried with various real life application and E3S benchmarks, showed better reduction in the communication energy as compared to other mapping algorithm such as crinkle and spiral. Since the number of tiles in the 3D NoC architecture is more so this architecture shows better results when tried with an application having greater number of the IP cores. Thus trying the design with random benchmark generated using TGFF [17] shows better result as compared to existing techniques. Compared to the Hybrid PSO and the proposed algorithm, has a better optimization effect when combined with various mapping algorithm.

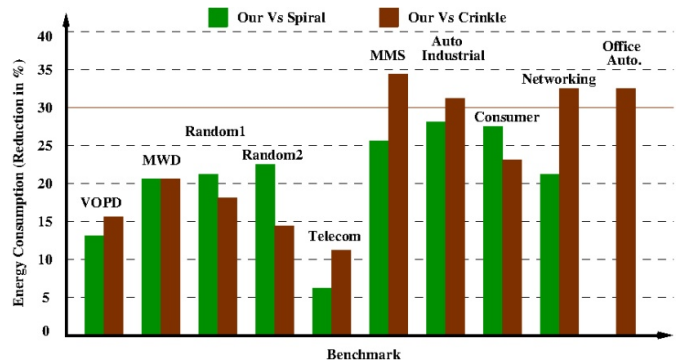

Figure 10. Average Latency reduction in % with various benchmarks.

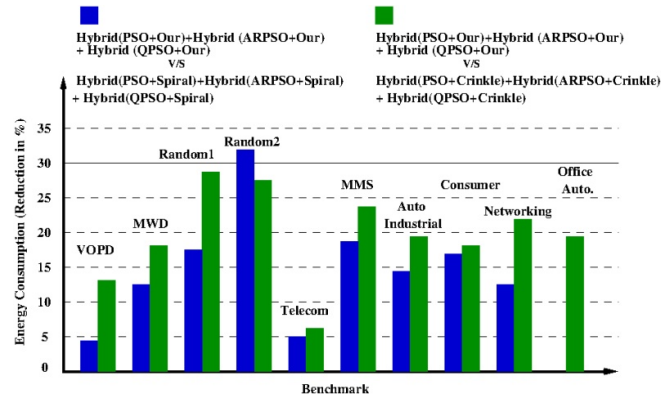

Figure 11. Average energy consumption reduction in % with various benchmarks in case of optimization algorithms.

## ACKNOWLEDGEMENT

I would like to thank "*ABV- Indian Institute of Information Technology & Management Gwalior*" for providing me the excellent research oriented environment.

**AUTHORS**

Vaibhav Jha has received his Master of Technology degree in specialization VLSI Design from Indian Institute of Information Technology and Management Gwalior in 2012. He has completed his Bachelors of Engineering degree in 2009. His academic research interest is in the area of Real time system, High Performance Computing, System on Chip, Distributed System, Computer Architecture, Databases, Networking and Communication.

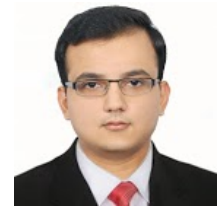

Sunny Deol, Master of Technology degree in Computer Science and Engineering from Indian Institute of Information Technology and Management Gwalior 2012. He has completed his Bachelor degree in Computer Science in 2010. Area of Interest includes Network on Chip, high speed network, Soft Computing, Embedded System and Computer Networks and Communication.

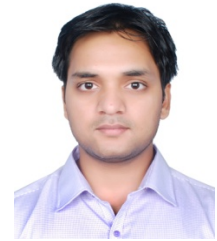